\documentclass[12pt]{article}

\newcommand{\sth}{\sigma^3}
\newcommand{\stw}{\sigma^2}

\newcommand{\sfo}{\sigma^4}
\newcommand{\alstwsfo}{\alpha_{\sigma^2\sigma^4}}
\newcommand{\Astwsfo}{A_{\sigma^2\sigma^4}}
\newcommand{\Astw}{A_{\sigma^2}}
\newcommand{\sumsfosupstw}{\sum_{\sigma^4 \supset \sigma^2}}
\newcommand{\sumstwsubsfo}{\sum_{\sigma^2 \subset \sigma^4}}
\newcommand{\sumstw}{\sum_{\sigma^2}}
\newcommand{\sumsfo}{\sum_{\sigma^4}}
\newcommand{\bisim}{\triangle\!\!\!\nabla}
\newcommand{\Nstw}{N_{\sigma^2}}
\newcommand{\Sbisimsfo}{S_{\triangle\!\!\!\nabla}(\sigma^4 )}
\newcommand{\Omsth}{\Omega_{\sigma^3}}
\newcommand{\Omsthsfo}{\Omega_{\sigma^3\sigma^4}}
\newcommand{\SSOthonbisim}{S_{\triangle\!\!\!\nabla}^{\rm SO(3,1)}}
\newcommand{\SSOthbisim}{S_{\triangle\!\!\!\nabla}^{\rm SO(3)}}
\newcommand{\SSUtwbisim}{S_{\triangle\!\!\!\nabla}^{\rm SU(2)}}
\newcommand{\vstwsfo}{v_{\sigma^2\sigma^4}}
\newcommand{\Rstwsfo}{R_{\sigma^2\sigma^4}}
\newcommand{\Arcsin}{{\rm Arcsin}}
\newcommand{\plS}{\,^+\!S}
\newcommand{\miS}{\,^-\!S}
\newcommand{\pmS}{\,^{\pm}\!S}
\newcommand{\pmSSUtwbisim}{\,^{\pm}\!S_{\triangle\!\!\!\nabla}^{\rm SU(2)}}
\newcommand{\pmSSOthbisim}{\,^{\pm}\!S_{\triangle\!\!\!\nabla}^{\rm SO(3)}}
\newcommand{\pmbvstwsfo}{\,^{\pm}\!\mbox{\boldmath$v$}_{\sigma^2\sigma^4}}
\newcommand{\pmvstwsfo}{\,^{\pm}\!v_{\sigma^2\sigma^4}}
\newcommand{\pmRstwsfo}{\,^{\pm}\!R_{\sigma^2\sigma^4}}
\newcommand{\pmbvstw}{\,^{\pm}\!\mbox{\boldmath$v$}_{\sigma^2}}
\newcommand{\pmbv}{\,^{\pm}\!\mbox{\boldmath$v$}}
\newcommand{\pmbw}{\,^{\pm}\!\mbox{\boldmath$w$}}
\newcommand{\pmbr}{\,^{\pm}\!\mbox{\boldmath$r$}}
\newcommand{\pmO}{\,^{\pm}\!\Omega}
\newcommand{\pmR}{\,^{\pm}\!R}
\newcommand{\alfofopl}{\alpha_{4(123)4^+}}
\newcommand{\algfopl}{\alpha_{\gamma (\alpha\beta 4)4^+}}
\newcommand{\etfofopl}{\eta_{4(123)4^+}}
\newcommand{\etgfopl}{\eta_{\gamma (\alpha\beta 4)4^+}}
\newcommand{\albg}{\alpha_{\beta (\alpha 44^+)\gamma}}
\newcommand{\Sbisim}{S_{\triangle\!\!\!\nabla}}
\newcommand{\pmSSOth}{\,^{\pm}\!S^{\rm SO(3)}}
\newcommand{\pmbvfoplfo}{\,^{\pm}\!\mbox{\boldmath$v$}_{(123)4^+4}}
\newcommand{\pmbvbg}{\,^{\pm}\!\mbox{\boldmath$v$}_{(\alpha 44^+)\beta\gamma}}
\newcommand{\pmbvfopla}{\,^{\pm}\!\mbox{\boldmath$v$}_{(\beta\gamma 4)4^+\alpha}}
\newcommand{\pmbvafo}{\,^{\pm}\!\mbox{\boldmath$v$}_{(\beta\gamma 4^+)\alpha 4}}

\newcommand{\pmRfoplfo}{\,^{\pm}\!R_{(123)4^+4}}
\newcommand{\pmRbg}{\,^{\pm}\!R_{(\alpha 44^+)\beta\gamma}}
\newcommand{\pmRfopla}{\,^{\pm}\!R_{(\beta\gamma 4)4^+\alpha}}
\newcommand{\pmRafo}{\,^{\pm}\!R_{(\beta\gamma 4^+)\alpha 4}}

\newcommand{\Na}{N_{(\alpha 44^+)}}

\newcommand{\Rfoplfo}{R_{(123)4^+4}}
\newcommand{\Rbg}{R_{(\alpha 44^+)\beta\gamma}}
\newcommand{\Rfopla}{R_{(\beta\gamma 4)4^+\alpha}}
\newcommand{\Rafo}{R_{(\beta\gamma 4^+)\alpha 4}}

\newcommand{\Omfo}{\Omega_{(1234^+)}}
\newcommand{\OmTfopl}{\Omega^{\rm T}_{(1234)}}
\newcommand{\OmTb}{\Omega^{\rm T}_{(\gamma\alpha 44^+)}}
\newcommand{\Omg}{\Omega_{(\alpha\beta 44^+)}}
\newcommand{\OmTa}{\Omega^{\rm T}_{(\beta\gamma 44^+)}}
\newcommand{\Oma}{\Omega_{(\beta\gamma 44^+)}}
\newcommand{\Omfopl}{\Omega_{(1234)}}

\renewcommand{\d}{{\rm d}}

\newcommand{\lfoplfo}{l_{4^+4}}
\newcommand{\lfoa}{l_{4 \alpha}}
\newcommand{\lfoon}{l_{41}}
\newcommand{\lfotw}{l_{42}}
\newcommand{\lfoth}{l_{43}}

\newcommand{\D}{{\cal D}}

\begin{document}

\title{Independent 4-tetrahedra connection representation of Regge calculus}
\author{V.M. Khatsymovsky \\
 {\em Budker Institute of Nuclear Physics} \\ {\em
 Novosibirsk,
 630090,
 Russia}
\\ {\em E-mail address: khatsym@inp.nsk.su}}
\date{}
\maketitle
\begin{abstract}
We consider simplest piecewise flat manifold consisting of two identical 4-tetrahedra (call it bisimplex). General relativity action for arbitrary piecewise flat manifold can be expressed in terms of sum of the (half of) bisimplex actions. We use representation of each bisimplex action in terms of certain rotation matrices (connections). This gives representation of any minisuperspace piecewise flat gravity system in terms of connections which do not connect neighboring 4-tetrahedra (more appropriate would be call these self-connections). If Regge calculus with independent 4-tetrahedra is considered, i. e. when the length of an edge is not constrained to be the same for all the 4-tetrahedra containing this edge, self-connection representation leaves 4-tetrahedra independent also in connection matrices sector. Action remains sum of independent 4-tetrahedra terms.
\end{abstract}

PACS numbers: 04.60.-m Quantum gravity

\newpage

Recently one often considers a modification of the genuine Regge calculus (RC) \cite{Regge} where the same edge can have different lengths depending on the 4-tetrahedron where it is defined, namely, the so-called area RC \cite{BarRocWil,RegWil} or simply RC with independent 4-tetrahedra \cite{Kha3}. If we additionally try to invoke description of the minisuperspace RC system in terms of tetrad and connection \cite{Fro}, the different 4-tetrahedra can not be treated as independent ones in the connection sector even if these are independent in the edge length sector. An idea is to apply connection representation separately to the (properly specified) contribution to the action of the different 4-tetrahedra.

We start with the standard Regge action \cite{Regge}
\begin{equation}\label{S/2}
{1\over 2}S = \sumstw \left ( 2\pi - \sumsfosupstw \alstwsfo \right ) \Astw.
\end{equation}

\noindent Here $\alstwsfo$ is hyperdihedral angle of the 4-simplex $\sfo$ at the 2-face $\stw$, $\Astw$ is the area (generally complex) of the triangle $\stw$.

Let us define for each 4-simplex $\sfo$ simplicial complex $\bisim$ built of only two identical, up to reflection w.r.t. any 3-face, 4-simplices one of which is just $\sfo$ and vertices of which are mutually identified. Call it here bisimplex. Its Regge action is
\begin{equation}
{1\over 2}S_{\bisim}(\sfo ) = \sumstwsubsfo (2\pi - 2\alstwsfo )\Astw.
\end{equation}

The action (\ref{S/2}) can be written as a sum over 4-simplices. When doing this, generalization to the case of independent 4-simplices is natural. Then, e.g., we should consider instead of $\Astw$ the set of values $\Astwsfo$ depending on the 4-simplex $\sfo \supset \stw$ where the area of $\stw$ is taken. Generalization to this case is not unique. The most symmetrical one (w.r.t. the different 4-simplices $\sfo \supset \stw$) reads
\begin{equation}\label{Sindep}
{1\over 2}S = \sumsfo \sumstwsubsfo \left ({2\pi\over \Nstw} - 2\alstwsfo \right )\Astwsfo,
\end{equation}

\noindent where $\Nstw$ is the number of the 4-simplices meeting at $\stw$. The terms in the sum over $\sfo$ depend on each other only through discrete value $\Nstw$; locally these are independent.

Important is that the full action is represented as sum over 4-simplices $\sfo$ of half of $\Sbisimsfo$ plus combination of areas,
\begin{equation}\label{S=Sbisim}
{1\over 2}S = \sumsfo \left [ {1\over 4}\Sbisimsfo + \sumstwsubsfo \left ({2\pi\over \Nstw} - \pi \right )\Astwsfo \right ].
\end{equation}

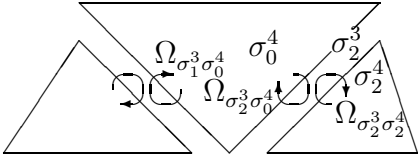
\begin{figure}
\setlength{\unitlength}{1.0mm}
\begin{picture}(55,20)
\put(0,0){\line(2,3){10}}

\put(0,0){\line(1,0){25}}
\put(10,15){\line(1,-1){15}}
\put(10,20){\line(1,-1){20}}
\put(10,20){\line(1,0){40}}
\put(30,0){\line(1,1){13.5}}
\put(50,20){\line(-1,-1){3.55}}

\put(16.5,8.5){\oval(4,4)[rt]}
\put(16.5,8.5){\oval(4,4)[rb]}
\put(16.5,8.5){\oval(4,4)[lt]}

\put(16.5,6.5){\vector(-1,0){1}}

\put(21.5,8.5){\oval(4,4)[lb]}
\put(21.5,8.5){\oval(4,4)[rb]}
\put(21.5,8.5){\oval(4,4)[lt]}

\put(21.5,10.5){\vector(1,0){1}}

\put(35,0){\line(1,0){20}}
\put(35,0){\line(1,1){15}}

\put(55,0){\line(-1,3){5}}

\put(38.5,8.5){\oval(4,4)[lb]}
\put(38.5,8.5){\oval(4,4)[rb]}
\put(38.5,8.5){\oval(4,4)[rt]}

\put(36.5,8.5){\vector(0,1){1}}

\put(43.5,8.5){\oval(4,4)[rt]}
\put(43.5,8.5){\oval(4,4)[lb]}
\put(43.5,8.5){\oval(4,4)[lt]}

\put(45.5,8.5){\vector(0,-1){1}}

\put(32.5,13){$\sigma^4_0$}

\put(20,12){$\Omega_{\sigma^3_1\sigma^4_0}$}
\put(26.5,7){$\Omega_{\sigma^3_2\sigma^4_0}$}
\put(44,4){$\Omega_{\sigma^3_2\sigma^4_2}$}

\put(43.5,13.5){$\sigma^3_2$}

\put(46.5,9){$\sigma^4_2$}

\end{picture}
\caption{To interpreting matrices $\Omsthsfo$.}\label{self-connection}
\end{figure}

\noindent We can use representation of the minisuperspace Regge action in terms of edge vectors and finite rotation SO(3,1) matrices \cite{Kha}. If we apply this immediately to $S$ on the independent 4-simplices, this independence no longer holds because rotation connection matrix $\Omsth$, $\sth = \sfo_1 \cap \sfo_2$, entering expressions for the defect angles refers to both 4-tetrahedra $\sfo_1, \sfo_2$ sharing the given 3-face $\sth$, not to a single one, and an expression for defect angle on $\stw$ refers to a set of such 3-faces $\sth \supset \stw$, i. e. to a set of the pairs $\sfo_1, \sfo_2$, $\sfo_1 \cap \sfo_2 = \sth$. The idea is to apply such representation to bisimplex actions in (\ref{S=Sbisim}). Instead of rotations with usual geometric interpretation (rotation between the local frames of neighboring $\sfo_1, \sfo_2$) we have matrices $\Omsthsfo$ which do not refer to any other 4-simplex than the given $\sfo$, see fig.\ref{self-connection}.

\noindent The bisimplex action has the form \cite{Kha,Kha2}
\begin{equation}
\SSOthonbisim (\sfo ) = \sumstwsubsfo \sqrt{\vstwsfo \circ \vstwsfo} \Arcsin {\vstwsfo \circ \Rstwsfo \over \sqrt{\vstwsfo \circ \vstwsfo}}.
\end{equation}

\noindent Here $\vstwsfo^{ab} = {1\over 2}\epsilon^{ab}{}_{cd}l^c_1l^d_2$ is bivector of the 2-face $\stw$ formed by the pair of the vectors $l^a_1, l^a_2$, $v\circ R \equiv {1\over 2}v_{ab} R^{ab}$, $\vstwsfo\circ \vstwsfo = 2\Astwsfo$. There are 5 connection matrices $\Omsthsfo$ and 10 curvature matrices $\Rstwsfo$, each $R$ being product of certain two matrices $\Omega^{\pm 1}$. The $\Arcsin$ means proper solution for the inverse function to $\sin$ while $\arcsin$ means principal value whose real part at real argument lays in the region $[-\pi/2, +\pi/2]$. To express action in terms of $\arcsin$ in appropriate physical sector is our particular task.

Also in \cite{Kha} we have suggested to use in the connection representation only self- or antiselfdual parts of the tensors $v$ and generators of $\Omega$ (and thus of $R$). That is, SU(2) rotations can be considered. In \cite{Kha2} we have used sum of self- and antiselfdual parts for analysis. Besides that, we have analyzed there representation with (anti-)selfdual parts described by SO(3) rotations, i. e. in the adjoint representation. The bisimplex action in these representations takes the form
\begin{eqnarray}
S & = & \plS + \miS, \label{S0+S0}\\
\pmSSUtwbisim & = & \sumstwsubsfo \sqrt{\pmbvstwsfo^2} \Arcsin {\pmvstwsfo \circ \pmRstwsfo \over \sqrt{\pmbvstwsfo^2}}, \\
\pmSSOthbisim & = & \sumstwsubsfo {1\over 2} \sqrt{\pmbvstwsfo^2} \Arcsin {\pmbvstwsfo * \pmRstwsfo \over \sqrt{\pmbvstwsfo^2}}.
\end{eqnarray}

\noindent Here $\pmvstwsfo$ are (anti-)selfdual parts of area tensors $\vstwsfo$. The $\pmbvstwsfo$ are 3-vectors parameterizing these parts, $\sqrt{\pmbvstw^2} = 2\Astw$, $v * R \equiv {1\over 2}v^a R^{bc} \epsilon_{abc}$. In Minkowsky spacetime $\pmbv$ and $\pmbw, \pmbr$ parameterizing generators of $\pmO, \pmR$ are generally complex, $\plS = (\miS)^*$. On equations of motion for $\Omega$'s (i. e. on-shell) $\plS$ and $\miS$ contribute the same half of action $S$. Therefore $S$ (\ref{S0+S0}) can be generalized by rescaling $\pmS$ by complex constants, $\plS \to C\plS$, $\miS \to C^*\miS$ so that their sum on-shell would result in Regge action: $C + C^* = 2$, $C = 1 + i/\gamma$,
\begin{equation}\label{S+S}
S \left (\equiv \sumsfo S(\sfo \right ) = \left (1 + {i\over \gamma}\right )\plS + \left (1 - {i\over \gamma}\right )\miS.
\end{equation}

\noindent Notation $\gamma$ is introduced here to provide analogy with Barbero-Immirzi parameter, coefficient at the term added to the Einstein action in the Cartan-Weyl form and vanishing on-shell (on equations of motion for connections)\cite{gamma}. Since, however, there is no direct correspondence (in, e. g., continuum limit) between our connection and Cartan-Weyl one, this analogy is purely formal. Also note that in the continuum theory (anti-)selfdual representation follows identically via decomposing full Cartan-Weyl 
action into self- and antiselfdual parts. Besides that, continuum analogs of SU(2) and SO(3) connection representations coincide (up to some redenoting). Contrary to that, all these representations in the considered exact discrete form are different (coincide only on-shell).

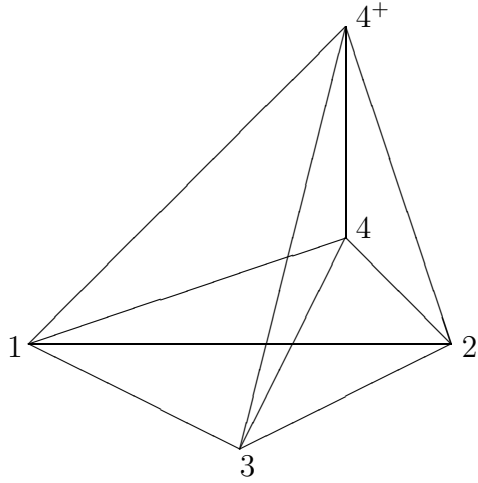
\begin{figure}
\setlength{\unitlength}{1pt}
\begin{picture}(200,200)(0,20)
\put (150,60){\line(1,1){120}}
\put (150,60){\line(3,1){120}} \put (150,60){\line(1,0){160}}
\put (150,60){\line(2,-1){80}}   
\put (270,100){\line(0,1){80}} \put (270,100){\line(1,-1){40}}
\put (270,180){\line(1,-3){40}}
\put (230,20){\line(1,4){40}}   
\put (230,20){\line(1,2){40}}   
\put (230,20){\line(2,1){80}}   
\put (270,180){$~4^{+}$} \put (142,55){$1$} \put
(270,100){$~4$} \put (310,55){$~2$} \put (230,10){$3$}
\end{picture}
\caption{Typical 4-simplex.}\label{4simplex}
\end{figure}

To pass to principal value of $\arcsin$, let us specify sector of variation of the angles. We adopt a regular way of constructing 4-dimensional simplicial complex from 3-dimensional analogous complexes ('leaves'). The typical 4-simplex is $(12344^+)$ with vertices $12344^+$ (fig.\ref{4simplex}).

\noindent These notations for vertices are the particular case of $i, k, l, \dots$ for vertices of certain 3-dimensional leaf. The $i^+$ and $i^-$ are future and past in time images of vertex $i$. The scheme of constructing 4-dimensional geometry is by successive shifting the vertices $i, k, l, \dots$ of current 3D leaf to $i^+, k^+, l^+, \dots$ of the 'next-in-time' leaf of the analogous structure (i. e. scheme of connection of the different vertices by links). When $i$ has evolved in time to $i^+$, $i^+$ is connected by {\it diagonal} links with those neighbors of $i$ in the leaf $k, l, m, \dots$ which are not yet evolved to the next leaf. Call the links $(ii^+)$, $(ii^-)$ {\it t-like} ones, to reserve 'timelike' for the local frame indices. The {\it leaf} links are completely contained in the leaf. We just get 4-simplices like $(12344^+)$ (and also $(123^+44^+)$, $(12^+3^+44^+)$, $(1^+2^+3^+44^+)$ and those with replacement $i^+ \to i$, $i \to i^-$).

It is natural to consider typical physical sector where t-like links are timelike, and leaf links are spacelike. Besides that, we adopt that diagonal links are also spacelike (that is, the distance between neighboring leaves, analog of lapse function, is sufficiently smaller than the typical linklength in the leaf). Thus, the only timelike link in the simplex $(12344^+)$ is $(44^+)$, others are spacelike ones.

This defines ranges for possible values of the dihedral angles. Denote by $\alfofopl$ the angle $\alstwsfo$ on $\stw = (123)$ in $\sfo = (12344^+)$ and analogously for others. It is not difficult to conclude that
\begin{equation}
\alfofopl = i\etfofopl
\end{equation}

\noindent as angle between 2 spacelike 3-faces (1234), $(1234^+)$ with opposite timelike link $(44^+)$. On other 6 spacelike triangles we find
\begin{equation}
\algfopl = {\pi \over 2} + i\etgfopl,
\end{equation}

\noindent $\alpha, \beta, \gamma, \dots = 1, 2, 3$ plus permutations $4 \leftrightarrow 4^+$. The $\eta$ is everywhere real. On the remaining 3 timelike triangles $(\alpha 44^+)$ we get real angles $\albg$ analogous to usual dihedral angles in the Euclidean geometry of 3D leaf.

Now express $\alstwsfo$ in (\ref{Sindep}) in terms of 'arcsin' functions , the same as in representations for $\Sbisim$. In $\SSOthonbisim$, $\SSOthbisim$ we might (modulo possible torsion) have on-shell
\begin{eqnarray}
\arcsin \sin (2\pi - 2\algfopl) & = & + 2i\etgfopl, \nonumber\\
\mbox{... the same for 4 $\leftrightarrow 4^+$ ...,} & & \nonumber\\
\arcsin \sin (2\pi - 2\alfofopl) & = & - 2i\etfofopl.
\end{eqnarray}

\noindent In $\SSUtwbisim$ each 'arcsin' gives only half of the corresponding angle appearing in $\SSOthonbisim$, $\SSOthbisim$, in particular,
\begin{eqnarray}
\arcsin \sin (\pi - \algfopl ) & = & {\pi\over 2} \pm i\etgfopl \\
\mbox{... the same for 4 $\leftrightarrow 4^+$ ...,}. & & \nonumber
\end{eqnarray}

\noindent Ambiguity arises because $\arcsin z$ is here on the cut $\Im z = 0, z^2 > 1$, where it undergoes discontinuity. To resolve the latter, one should add certain $\pm i0$ to $z$. Here, however, we choose to limit ourselves by SO(3,1), SO(3) representations, or, for calculational simplicity, by SO(3) one.

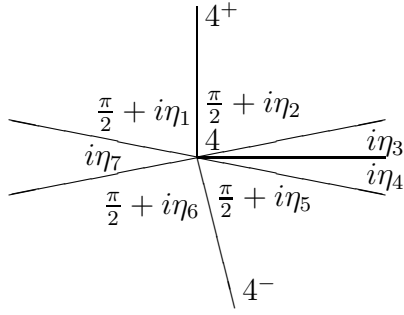
\begin{figure}
\setlength{\unitlength}{0.5mm}
\begin{picture}(100,120)(-10,-10)
\put (60,50){\line(0,1){40}}
\put (60,50){\line(1,-4){10}}
\put (60,50){\line(5,-1){50}}
\put (60,50){\line(5,0){50}}
\put (60,50){\line(5,1){50}}
\put (60,50){\line(-5,1){50}}
\put (60,50){\line(-5,-1){50}}
\put (62,85){$4^{+}$}
\put (62,52){$4$}
\put (72,12){$4^{-}$}
\put (62,62){${\pi \over 2} + i\eta_2$}
\put (65,37){${\pi \over 2} + i\eta_5$}
\put (35,35){${\pi \over 2} + i\eta_6$}
\put (33,60){${\pi \over 2} + i\eta_1$}
\put (105,52){$i\eta_3$}
\put (105,43){$i\eta_4$}
\put (30,48){$i\eta_7$}
\end{picture}
\caption{Dihedral angles on the leaf/diagonal triangle.}\label{angles}
\end{figure}

The above angles of the type $i\eta$ and $\pi /2 + i\eta$ define defect angle on the leaf or diagonal triangles. For our way of constructing 4-dimensional geometry, there are 4 angles between the t-like and leaf/diagonal 3-faces of the type $\pi /2 + i\eta$ and a few (from 0 to 4) angles of the type $i\eta$ (fig.\ref{angles}).

\noindent As a result, proportional to $\pi$ contribution to defect angle is canceled and imaginary value remains.

Finally, for 3 real angles we adopt the sector $\pi /4 < \albg < 3\pi /4$ in which
\begin{equation}
\arcsin \sin (2\pi - 2 \albg ) = 2\albg - \pi.
\end{equation}

As a result, we have
\begin{eqnarray}
 & & \pmS = \pmSSOth = {1\over 2} \sumsfo \left \{{1\over 2}\sqrt{\pmbvfoplfo^2}\arcsin {\pmbvfoplfo * \pmRfoplfo\over \sqrt{\pmbvfoplfo^2}} \right. \phantom{ = \equiv \sumsfo \pmS (\sfo )} \nonumber \\ & & + \sum_{\stackrel{(\alpha,\beta,\gamma)={\rm cycle}}{{\rm perm}~(1,2,3)}}\left [ \sqrt{\pmbvbg^2} \left ({2\pi\over \Na} - {\pi\over 2}\right ) \right. \nonumber \\ & & \phantom{+ \sum_{\stackrel{(\alpha,\beta,\gamma)={\rm cycle}}{{\rm perm}~(1,2,3)}}} - {1\over 2} \sqrt{\pmbvbg^2} \arcsin {\pmbvbg * \pmRbg \over \sqrt{\pmbvbg^2}} \nonumber \\ & & \phantom{+ \sum_{\stackrel{(\alpha,\beta,\gamma)={\rm cycle}}{{\rm perm}~(1,2,3)}}} - {1\over 2} \sqrt{\pmbvafo^2} \arcsin {\pmbvafo * \pmRafo \over \sqrt{\pmbvafo^2}} \nonumber \\ & & \phantom{+ \sum_{\stackrel{(\alpha,\beta,\gamma)={\rm cycle}}{{\rm perm}~(1,2,3)}}} \left. \left. - {1\over 2} \sqrt{\pmbvfopla^2} \arcsin {\pmbvfopla * \pmRfopla \over \sqrt{\pmbvfopla^2}} \right ] \right \} \equiv \sumsfo \pmS (\sfo )
\end{eqnarray}

\noindent (to be substituted to (\ref{S+S})). Explicit expressions of $R$ in terms of $\Omega$ are, e. g.,
\begin{eqnarray}
\Rfoplfo = \OmTfopl \Omfo, \nonumber \\
\Rbg = \OmTb \Omg, \nonumber \\
\Rafo = \OmTa \Omfo, \nonumber \\
\Rfopla = \OmTfopl \Oma, \\
(\alpha,\beta,\gamma ) = \mbox{cycle perm} (1,2,3). \nonumber
\end{eqnarray}

\noindent Note that purely area terms in $S$ are only 3 t-like areas $\sqrt{\pmbvbg^2}$, in some sense gauge ones. These might parameterize 3 components of the vector of the link $(44^+)$, a discrete analog of the lapse-shift vector of general relativity.

To resume, we have considered connection representation for the minisuperspace RC gravity action with independent 4-tetrahedra leaving 4-tetrahedra independent. Requirement has been studied and satisfied that on-shell connections resulting in genuine Regge action in Minkowsky spacetime might not coincide with points of non-analyticity. Because of independence of the 4-tetrahedra path integral measure should factorize over 4-tetrahedra. Since separate 4-tetrahedron possesses very few number of the degrees of freedom as compared to real gravity system, the measure on it is very simple and follows from symmetry considerations and correspondence with continuum case. In the latter the local measure at a point is defined up to scalar density for which different powers of $\det \|g_{\lambda\mu}\|$ were considered \cite{Mis,DeW}. Also factor $g^{00}$ could be inserted \cite{Leu,FraVil}.
\begin{eqnarray}
\d \mu (\sfo ) = e^{iS(\sfo )} (\epsilon_{abcd}\lfoplfo^a \lfoon^b \lfotw^c \lfoth^d)^{\zeta_1}[(\epsilon_{abcd} \lfoon^b \lfotw^c \lfoth^d)^2]^{\zeta_2} \d^4 \lfoplfo \prod^3_{\alpha = 1} \d^4\lfoa \nonumber \\ \cdot \D \Omfopl \D \Omfo \prod_{\stackrel{(\alpha,\beta,\gamma)={\rm cycle}}{{\rm perm}~(1,2,3)}} \D \Oma. \phantom{\d^4 \lfoplfo \prod^3_{\alpha = 1} \d^4\lfoa }
\end{eqnarray}

\noindent The $\lfoplfo^a, \lfoa^a, \alpha = 1,2,3$ is tetrad of independent edge vectors, $\zeta_1$, $\zeta_2$ are parameters, $\zeta_2$ = 0 or 1, $\D \Omega$ is SO(3,1) Haar measure. Prototypes of $\det \|g_{\lambda\mu}\|$ and of $g^{00} \det \|g_{\lambda\mu}\|$ are just 4- and 3-volumes squared, $(\epsilon_{abcd}\lfoplfo^a \lfoon^b \lfotw^c \lfoth^d)^2$ and $(\epsilon_{abcd} \lfoon^b \lfotw^c \lfoth^d)^2$, respectively. Integration over any one of five $\Omega$'s decouples by SO(3,1) symmetry, and that one over remaining $\Omega$'s reduces to integration over 4 independent $R$'s, e. g. over $\Rfoplfo, \Rfopla, (\alpha,\beta,\gamma ) = \mbox{cycle perm} (1,2,3)$. Together with the result of our previous papers \cite{Kha3} reducing the measure on the independent 4-tetrahedra to the measure in genuine RC, the latter is thus fixed.

The present work was supported in part by the Russian Foundation for Basic Research
through Grant No. 08-02-00960-a.



\begin{thebibliography}{99}
\bibitem{Regge}
 T. Regge, General relativity theory without coordinates. - Nuovo Cimento {\bf 19},
 568 (1961).
\bibitem{BarRocWil}
 J.W. Barrett, M. Ro\v{c}ek, R.M. Williams, A note on area variables in Regge
 calculus. - Class. Quantum Grav. {\bf 16}, 1373 (1999), gr-qc/9710056.
\bibitem{RegWil}
 T. Regge, R.M. Williams, Discrete structures in gravity. - Journ. Math. Phys.
 {\bf 41}, 3964 (2000), gr-qc/0012035.
\bibitem{Kha3}
 V.M.Khatsymovsky, Regge calculus from discontinuous metrics. - Phys. Lett.
 {\bf 567B}, 288 (2003), gr-qc/0304006.\\
 V.M. Khatsymovsky, Gravity action on discontinuous metrics, arXiv:0808.xxxx [gr-qc] (2008).
\bibitem{Fro}
 J. Fr\"{o}hlich, Regge Calculus and Discretized Gravitational Functional
 Integrals, I.~H.~E.~S.~preprint (1981) (unpublished);
 Non-Perturbative Quantum Field
 Theory: Mathematical Aspects and Applications, Selected Papers - World Scientific,
 Singapore, 523 (1992).
\bibitem{Kha}
 V.M. Khatsymovsky, Tetrad and self-dual formulations of Regge calculus. - Class. Quantum Grav. {\bf 6}, L249 (1989).
\bibitem{Kha2}
 V.M. Khatsymovsky, Feynman path integral in area tensor Regge calculus and positivity. - Phys. Lett. {\bf 601B}, 229 (2004), gr-qc/0406050.
\bibitem{gamma}
 S. Holst, Barbero's Hamiltonian Derived from a Generalized Hilbert-Palatini Action. - Phys. Rev. D, {\bf 53}, 5966 (1996), gr-qc/9511026.\\
 J.F. Barbero, Real Ashtekar Variables for Lorentzian Signature Space-times. - Phys.
 Rev. D {\bf 51}, 5507 (1995), gr-qc/9410014.\\
 G. Immirzi, Quantum Gravity and Regge Calculus. - Class. Quantum Grav. {\bf 14},
 L177 (1997), gr-qc/9701052.
\bibitem{Mis}
 C.W. Misner, Feynman quantization of general relativity. - Rev. Mod. Phys. {\bf 29}, 497 (1957).
\bibitem{DeW}
 B.S. DeWitt, Quantization of fields with infinite-dimensional invariance groups. III. Generalized Shwinger-Feynman theory. - Journ. Math. Phys. {\bf 3}, 1073 (1962).
\bibitem{Leu}
 H. Leutwyler, Gravitational field: equivalence of Feynman quantization and canonical
 quantization. - Phys. Rev. {\bf 134}, 1155 (1964).
\bibitem{FraVil}
 E.S. Fradkin, G.A. Vilkovisky, S matrix for gravitational field. II. Local measure;
 general relations; elements of renormalization theory. - Phys.Rev. D {\bf 8}, 4241 (1974).
\end{thebibliography}
\end{document}